\documentclass[aps,prl,twocolumn,a4paper,10pt]{revtex4-1}

\usepackage[T1]{fontenc}		%%% T1 Font Encoding
\usepackage[english]{babel}		%%% LaTeX for English
\usepackage[latin1]{inputenc}	%%% Accentuated Fonts
\usepackage{times}

\usepackage{amsmath}			%%% Mathematics AMS package
\usepackage{amssymb}			%%% Mathematics AMS package
\usepackage{bbm}				%%% Font for ensembes

\usepackage[colorlinks=true, a4paper=true, pdfstartview=FitV,linkcolor=blue, citecolor=blue, urlcolor=blue]{hyperref}

\usepackage{graphicx}			%%% Include graphics
\usepackage{graphics}			%%% Include graphics
\DeclareGraphicsExtensions{.pdf}
\usepackage{color}		

\newcommand{\bea}{\begin{eqnarray}}
\newcommand{\eea}{\end{eqnarray}}

\newcommand{\bk}{\mathbf{k}}
\newcommand{\bq}{\mathbf{q}}
\newcommand{\bsigma}{\boldsymbol{\sigma}}

\newcommand{\bx}{\mathbf{x}}

\begin{document}

\title{Strongly interacting Weyl semimetals: Stability of the semimetallic phase and \\ emergence of almost free fermions}

\author{Johan~Carlstr\"om and Emil~J. Bergholtz}
\affiliation{Department of Physics, Stockholm University, 106 91 Stockholm, Sweden}
\date{\today}

\begin{abstract}
Using a combination of analytical arguments and state-of-the-art diagrammatic Monte Carlo simulations we show that 
the corrections to the dispersion in interacting Weyl semimetals are determined by the ultraviolet cutoff and the inverse screening length. If both of these are finite, then the diagrammatic series is convergent even in the low-temperature limit, which implies that the semimetallic phase remains stable. Meanwhile,
the absence of a UV cutoff or screening results in logarithmic divergences at zero temperature. 
These results highlight the crucial impact of Coulomb interactions and screening, mediated e.g. through the presence of parasitic bands, which are ubiquitous effects in real-world materials.
Also, despite sizable corrections from Coulomb forces, the contribution from the frequency dependent part of the self energy remains extremely small, thus giving rise to a system of effectively almost free fermions with a strongly renormalized dispersion.

\end{abstract}
\maketitle

%%%%%%%%%%% INTRO %%%%%%%%%%%%

{\it Introduction.---}
Weyl semimetals (WSM's) display a band-structure that is gapped everywhere in the Brillouin zone, 
except at a set of nodal points where the valence and conduction bands meet, leading to an effective low-energy description in terms of massless  Weyl fermions with a linear spectrum and thus parallels in relativistic physics \cite{RevModPhys.90.015001,volovik2009universe,PhysRevB.83.205101,1367-2630-9-9-356,HOSUR2013857}. 
While often compared to graphene, the Weyl semimetals are distinct from other Dirac materials in that the nodal structure of their electronic spectrum is topologically protected, and cannot be gapped by perturbations. 
The origin of this band structure is closely related to explicit breaking of inversion or time-reversal symmetry  \cite{RevModPhys.90.015001}. Thus, TaAs, which was the first material identified as a WSM \cite{doi:10.1146/annurev-conmatphys-031016-025225} exhibits a non-centrosymmetric crystal structure \cite{PhysRevX.5.011029}, and preceding proposals for experimental realizations include adding symmetry breaking elements to artificially constructed Dirac materials \cite{PhysRevB.84.235126}. 

The distinct spectrum, topology and broken symmetries of WSMs give rise to a number of exotic electronic properties without parallel in other systems. 
A prominent example, considered long before the realization of these physics in condensed matter systems, is the chiral anomaly, which is a magnetic response where the nodal points act as sources and sinks of a spontaneous current. Notably, this property was exploited as an experimental signature in TaAs \cite{Xu613}.
This work also found evidence of Fermi arcs, which is a type of surface state of topological origin that is presently attracting broad interest \cite{RevModPhys.90.015001,doi:10.1146/annurev-conmatphys-031016-025225}.

%A more recent finding which is currently not well understood but with potential technological applications is the titanic magnetoresistance (the largest currently known), which occurs in the overtilted, and thus metallic system WTe$_2$ \cite{10.1038/nature13763,10.1038/nature15768}. 
%%
%Thus, the topological band structure and broken symmetries are conspicuously manifested in a range of exotic electronic properties.

%10.1038/nature13763 10.1038/nature15768,
Although WSM's have attracted wide interest from theorists, the role of interactions in these systems has not been extensively studied. Most likely this is rooted in the fact that much of the novel physics in these materials emerges already at the level of a noninteracting description, but also that the topological nature of the band structure seems to limit the scope of interactions.  
 A natural comparison for the WSM's is graphene, where interactions are known to renormalize the Fermi velocities, and this apparently raises questions about the interplay of Coulomb repulsion with anisotropies and tilting of the dispersion, which are particularly prominent features of Weyl semimetals due to their symmetry breaking nature  \cite{PhysRevB.91.115135,10.1038/nature15768}. The current understanding of this problem is based on renormalisation group studies that indicate flow towards an untilted system in the low energy limit \cite{PhysRevB.96.195157}.

A few previous works have touched upon some of the fundamental properties of interacting Weyl or Dirac materials.
In Ref.  \cite{PhysRevLett.113.136402} it is argued that the characterization of the interacting system in terms of topological charge remains meaningful in the interacting regime and that the many-body Hamiltonian can be adiabatically connected to a topological Hamiltonian which takes into account frequency-independent corrections, and thus can be understood as a renormalized dispersion. 
In Refs. \cite{10.1038/srep19853, 2018arXiv180405078M}, on the other hand, it was shown that WSM's can be gapped by the inclusion of interactions that are local in $k$-space, though such inter-particle forces are unlikely to be representative of realistic materials. 
In semimetallic lattice models a gap can also be opened by strong contact interactions, resulting in a Mott insulator, a scenario that may also occur for sufficiently strong Coulomb forces \cite{Araki}. Variational cluster calculations furthermore indicate charge and spin density wave instabilities in this scenario \cite{PhysRevB.94.241102}.
However before the onset of Mott physics when the system is still gapless, RPA calculations indicate marginal Fermi liquid behavior \cite{PhysRevB.92.045104}.

Interactions in semimetallic systems are in general a rather subtle topic from a methodological point of view. The vanishing density of states at the Fermi level seems to make them ideal candidates for diagrammatic techniques, as the weight of higher order graphs is dramatically reduced when the system is gapped in almost the entire Brillouin zone. In addition to this, the nodal points are in many cases protected by symmetry \cite{PhysRevB.97.161102}. 
This picture is further corroborated by a theorem which states that the diagrammatic expansion for graphene remains analytic down to zero temperature given that the interactions are sufficiently weak and short ranged \cite{PhysRevB.79.201403}. 
At the same time, the vanishing density of states means that the system is essentially unscreened, which in graphene leads to logarithmically divergent corrections to the Fermi velocity \cite{PhysRevLett.118.026403,RevModPhys.84.1067}.

In this work we focus on the case of a single node with a linear spectrum and examine the premises for a convergent diagrammatic expansion using scaling transformation as well as diagrammatic Monte Carlo simulations, which allow us to stochastically sample the expansion.

%%%%%%%%%% DIAGRAMMATICS %%%%%%%%%%%%%%%%%
{\it Diagrammatic expansion.---} %analysis of the full Greens function}
In an isotropic and untilted Dirac system, the low-energy part of the spectrum can be modelled with a linear dispersion on the form 
\bea
H_0 =\Psi_\bk^\dagger v_f^0 \bsigma \cdot \bk \Psi_\bk,\label{H0}
\eea
where $v_f^0$ is the Fermi velocity, $\hbar=1$, and  $\Psi^\dagger_\bk, \Psi_\bk$ are two-component field operators. 
In the presence of two-particle interactions we obtain a bi-quadratic term in the Hamiltonian on the form
\bea
H_1=\int d\bx_1d\bx_2 \hat{n}_\alpha(\bx_1) V_{\alpha\beta}(\bx_1-\bx_2) \hat{n}_\beta(\bx_2),
\eea
where $\hat{n}_\alpha(\bx)=\psi_\alpha^\dagger(\bx)\psi_\alpha(\bx)$ is a density operator for the component $\alpha$. 
 
The Greens function for the interacting system can be obtained through an expansion in the bi-quadratic part of the Hamiltonian,
\bea\nonumber
G_{\alpha\beta}(\tau_a-\tau_b, \mathbf x_a-\mathbf x_b)=
Z^{-1}\sum_n \frac{(-1)^n}{n!} \int_0^\beta d \tau_i \; 
\\
\text{Tr}\{e^{-\beta H_0 } T[H_1(\tau_1)...H_1(\tau_n) \psi^\dagger_\alpha(\tau_a,\mathbf x_a)\psi_\beta(\tau_b,\mathbf x_b)] \}\label{expansion}
\eea
which in turn can be expressed as a set of connected diagrams via Wicks theorem \cite{fetter}. 

We can define a sub-class of topologies that contain tadpole insertions on the form 
\bea
V_{\alpha\beta}(\bk=0) G_{\alpha\alpha}(\tau\to 0^-,\bk) \Psi^\dagger_\beta(\tau,\mathbf x) \Psi_\beta(\tau,\mathbf x),
\eea
that can be equated to a shift of the chemical potential by $\sim V(\bk=0) \langle \hat{n}\rangle$.
Since we are interested in the system at a specific stoichiometry we proceed to drop these terms to obtain  
\bea
H_1'=H_1-H_{\text{tadpole}}.
\eea
From the point of view of a diagrammatic treatment, this is equivalent to rejecting contributions from topologies where a boson is emitted without being reabsorbed. 

A crucial property of the dispersion Eq. (\ref{H0}) is that it is odd under inversion, which in turn implies that central properties of the band structure are protected by symmetry.  In particular, it can be shown that the Greens function of the interacting system always has a pole in $\omega=0,\bk=0$, so that the nodal point is stable against interactions and the system remains gapless as long as the series is convergent. This result also holds for systems with anisotropies and broken particle-hole symmetry, as well as for certain lattice models, notably that of graphene
 \cite{PhysRevB.97.161102}. 

 %%%%%%%%%% SCALE TRANSFORM %%%%%%%%%%%%%%%%% 
{\it Scale transformation.---} 
In ideal Dirac systems with a linear dispersion we find a close relationship between temperature and scale, which can be exploited to formulate criteria for convergence of the series in the zero-temperature limit, and consequently, the stability of the semimetallic phase in the ground state. 
This becomes particularly transparent if we choose an energy scale such that the temperature is unity. As the partition function only depends on the  
relative scales of energy and temperature, i.e.,
\bea
z(\beta,H)=z(1, \beta H),
\eea
we can in principle conduct an expansion in a bare Greens function of the form
\bea
G_0(\omega,\bk,\beta)=\frac{1}{i\omega -\beta H_0(\bk)}=G_0(\omega, \beta\bk)
\eea
where $\omega=(2n+1)\pi$. Here, we have used linearity of $H_0(\bk)$ in the second equality. The corrections to the full Greens function then take the form
\bea
\!\delta G(\tau,\!\bk,\beta)\sim\prod_i^N\! d \bk_i \Big\{\! \prod_j^N \! \beta V(\bk_j)\prod_l^{2N+1}\! G^0_{\alpha_l\beta_l}(\tau_l,\! \beta\bk_l)\! \Big\}. \;\;\; 
\eea
Keeping $\bk$ fixed, we have $N$ independent momenta $\{\bk_i\}$ to integrate over, while $\{\bk_j\},\;\{\bk_l\}$ are linear functions of $\{\bk_i\}$, whose form depend on the diagram topology. We now introduce a scale transformation on the form $\bk'=\gamma \bk$ to obtain
\bea\nonumber
\delta G(\tau,\bk,\beta)\sim\prod_i^N  \frac{d\bk_i}{d\bk_i'}d\bk_i'\\\nonumber
 \times\Big\{ \prod_j^N \beta V(\bk_j'/\gamma)\;\prod_l^{2N+1} G^0_{\alpha_l\beta_l}(\tau_l, \beta\bk_l'/\gamma)\Big\}\\
=  \prod_i^N  d\bk_i' \;\prod_j^N \frac{\beta}{\gamma^D} V\Big[\frac{\bk_j'}{\gamma}\Big]\;\prod_l^{2N+1} G^0_{\alpha_l\beta_l}\Big[\tau_l, \beta\frac{\bk_l'}{\gamma}\Big]
\eea
Choosing $\gamma=\beta$ we get
\bea
\prod_i^N  d\bk_i' \;\prod_j^N \beta^{1-D} V\Big[\frac{\bk_j'}{\beta}\Big]\;\prod_l^{2N+1} G^0_{\alpha_l\beta_l}\Big[\tau_l, \bk_l'\Big]. \label{scaledG}
\eea
Thus, we have chosen our units such that the temperature dependence of the bare Green's function only enters implicitly through the scale. From Eq. (\ref{scaledG}) it is clear that the interaction terms scale according to
\bea
\beta V(\bk)\to \beta^{1-D}V(\bk'/\beta).\label{scaleV}
\eea
For an unscreened Coulomb interaction in 3D we find
\bea
\beta^{1-D}\frac{4\pi \alpha }{(\bk'/\beta)^2}=\frac{4\pi \alpha}{\bk'^2}
\eea
which is also only implicitly dependent on temperature. Thus, in this scenario the interacting theory is invariant under a group of conjugate scale- and temperature-changes so that formally
\bea
G(\tau,\beta,\bk)=G(\tau,1,\beta\bk).
\eea
It should be stressed however, that unscreened Coulomb repulsion gives a logarithmically divergent contribution to the Fermi velocity, as will be discussed below. 
If screening is present we obtain 
\bea
\beta^{1-D}\frac{4\pi \alpha }{(\bk'/\beta)^2+\lambda^{-2}_0}=\frac{4\pi \alpha}{\bk'^2+\beta^2 \lambda_0^{-2}}\label{screnedCoulomb}
\eea
implying that the screening length decreases as the temperature is lowered (and the scale changes). Likewise, if a UV cutoff is present, then it will behave as $\Lambda\sim\beta$, giving a scaling relation for the Greens function of the interacting system on the form
\bea
G(\tau,\beta,\bk,\lambda^{-1}_0, \Lambda_0)=G\big(\tau,1,\beta\bk,\beta \lambda^{-1}_0 ,\beta \Lambda_0 
\big)	\label{Grenorm}
\eea 
Thus, writing our theory on this form, the temperature effectively only enters in the form of a scale change. Moreover, the zero temperature limit is associated with the simultaneous divergence of the UV cutoff and the inverse screening length. Correspondingly, the ground state should be determined by the ratio $\Lambda_0/\lambda^{-1}_0$ provided that the series is convergent. 
This leads to two basic scenarios: Either both screening and a UV cutoff are present, in which case $\Lambda_0/\lambda^{-1}_0$ takes a finite value, or the ratio is infinite.  
Moreover, we find only one more (inverse) length scale in this problem, namely the Fermi velocity, which is independent of $\beta$. Thus, we should expect that the low temperature regime corresponds to $v_f^0 \lambda^{-1}\gg 1$ when screening is present.

%%%%%%%%%% UV AND IR %%%%%%%%%%%%%%%%%
{\it Ultraviolett physics.---}
From the scaling properties of the Greens function (Eq. \ref{Grenorm}) it appears that ultraviolet behavior is central to the convergence of the diagrammatic expansion at low temperatures, even in the presence of a cutoff. 

At the lowest order (Fock diagram) we obtain a self energy on the form
\bea
\Sigma(\bk)\!=\!\int \!\frac{d\bq}{(2\pi)^D} \frac{V(\bk-\bq,\beta)}{2} \big[G_0(\delta,\bq)+G_0(-\delta,\bq)\big].\;\;\;\;\;
\eea
Integrating over a shell of momenta $\beta \Lambda_0>|\bq|>k_0$ where $k_0\gg |\bk|$ we can expand the interaction in $\bk$ to obtain
\bea
\frac{4\pi\alpha}{(\bq-\bk)^2+\beta^2\lambda^{-2}_0}\approx \frac{4\pi\alpha}{\bq^2+\lambda^{-2}_0\beta^2}+\frac{8\pi\alpha\bq\cdot\bk }{(\bq^2+\lambda^{-2}_0\beta^2)^2}\;\;\;\;\;
\eea
where only the second term contributes to the integral. If $v_f^0 |\bq|\gg 1$, then we find  $[G_0(\delta,\bq)+G_0(-\delta,\bq)]\approx -\hat{\bq}\cdot\bsigma$ where $\hat{\bq}=\bq/|\bq|$. Taking $\bk=k e_z$ this gives
\bea\nonumber
\delta\Sigma(\bk)\sim\int_{k_0}^{\beta \Lambda_0}dq d\phi d\theta \sin\theta q^2 \frac{qk \cos \theta\;\hat{\bq}\cdot\bsigma}{(q^2+\lambda^{-2}_0\beta^2)^2}\\\nonumber
\sim\int d\phi d\theta \sin\theta   \cos \theta\;k \hat{\bq}\cdot\bsigma
\\
\times\Big[  \frac{\lambda^{-2}_0\beta^2}{q^2+\lambda^{-2}\beta^2}+\log[\lambda^{-2}_0\beta^2+q^2]  \Big]_{k_0}^{\beta \Lambda_0}.\label{UVdiv}
\eea 
In the low-temperature limit we find
\bea\nonumber
\lim_{\beta\to\infty}
 \Big[  \frac{\lambda^{-2}_0\beta^2}{q^2+\lambda^{-2}_0\beta^2}  \Big]_{k_0}^{\beta \Lambda_0}
+\log \Big[ \frac{\beta^2\lambda^{-2}_0+\beta^2 \Lambda_0^2}{\beta^2\lambda^{-2}_0+k_0^2}\Big]  
\\=\frac{\lambda^{-2}_0}{\Lambda_0^2+\lambda^{-2}_0}-1+\log[1+\Lambda_0^2\lambda^{2}_0]. \label{Fock}
\eea 
Taking the limit $\Lambda_0/\lambda_0^{-1}\to\infty$ this grows logarithmically. We are thus in a position to formulate a basic conjecture about the ground state behavior of this systems: Firstly, as explained above,   
we expect that the low temperature regime corresponds to 
\bea
v^0_f \lambda^{-1}\gg 1,\label{conj1}
\eea
and secondly, we anticipate that the corrections to the particle energy grow as 
\bea
\lim_{\Lambda_0/\lambda_0^{-1}\to\infty}\Sigma(\omega=0) \sim\log[\Lambda_0/\lambda_0^{-1}] .\label{conj2}
\eea

\begin{figure}[!htb]
\includegraphics[width=\linewidth]{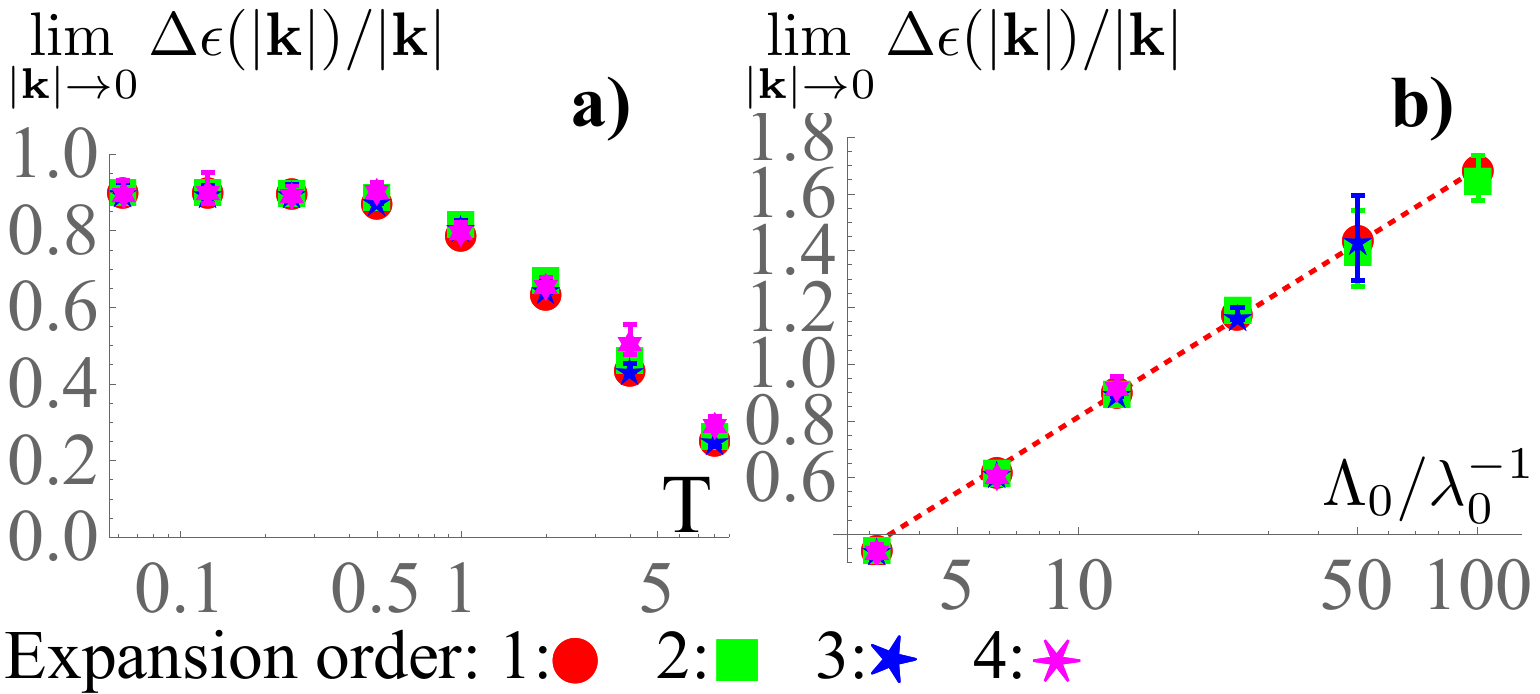}
\caption{
a) Frequency independent correction to the particle energy $\Delta \epsilon(\bk)$ (defined in Eq. \ref{sigmaComp}) to order $O=1,2,3,4$, near the nodal point as a function of temperature. 
The model parameters are given by $\Lambda_0=4\pi,\;\lambda^{-1}_0=1/4, \alpha=2, \;v_f^0 =4$, (see Eqs. \ref{H0},\;\ref{screnedCoulomb} for definitions).
b) Corrections to the dispersion as a function of $\Lambda_0/\lambda_0^{-1}$. Consistent with Eq. (\ref{conj2}), we find a straight line 
when the data is plotted on a logarithmic scale. The dashed line shows a fitted function on the form $f(\Lambda_0/\lambda_0^{-1})=a+b\log(\Lambda_0/\lambda_0^{-1})$. 
}
\label{Sigma_0}
\end{figure}

%%%%%%%%%% MONTE CARLO %%%%%%%%%%%%%%%%%
{\it Diagrammatic Monte Carlo simulation.---}
To test the conjecture outlined above, we employ diagrammatic Monte Carlo, which is a numerical protocol for stochastic sampling of the diagrammatic expansion \cite{PhysRevLett.99.250201}.
Within this framework, the space of connected diagram topologies for the self energy is sampled via a Metropolis type random walk. 
The Greens function is then obtained via Dyson's equation:
\bea
G(\omega,\bk)=\frac{1}{i\omega - H_0(\bk)-\Sigma(\omega,\bk)}.\label{Dyson}
\eea
The sampling protocol used in this work is based on the worm algorithm, and is described in \cite{PhysRevB.97.075119}. A central point of this approach is that {\it all} diagram topologies are sampled up to a given expansion order $N$. Thus, the only systematic source of error is truncation of the series, meaning that if it converges, it does so to a result that is unbiased and controllable. 

In principle we can divide the self energy into two parts according to
 \bea
 \Sigma(\omega,\bk)=\Delta \epsilon(\bk)+\tilde{\Sigma}(\omega,\bk),\label{sigmaComp}
 \eea
 where, $\Delta \epsilon(\bk)$, merely gives a correction to the kinetic energy with no frequency dependence. Thus, $H_0(\bk)+\Delta \epsilon(\bk)$ essentially describes a system of free fermions with a renormalised dispersion, while $\tilde{\Sigma}(\omega,\bk)$ encodes many-body effects  beyond the band-theory paradigm. 
 
 In Fig. (\ref{Sigma_0},a) the correction $\Delta \epsilon(\bk)$
 is displayed as a function of temperature for a system with a finite UV cutoff and screening. Firstly, we find that the convergence is very rapid with respect to expansion order, with corrections from higher orders terms being much smaller than those of the Fock term ($O=1$). Secondly, in agreement with Eq. (\ref{conj1}) we find that the temperature dependence saturates once the screening becomes the shortest length scale in the problem, which here corresponds to $T<1$. Indeed, at the lowest temperatures the differences are zero within the error-bars and we conclude that the ground state is a well defined limit of diagrammatic expansion. Furthermore, we recall the result \cite{PhysRevB.97.161102}, which states that if the series is convergent, then the semimetallic state is protected by symmetry. 
 
 To examine the conjecture Eq. (\ref{conj2}), we scale $\Lambda_0/\lambda_0^{-1}$ in the low temperature region where Eq. (\ref{conj1}) holds, see Fig. (\ref{Sigma_0},b). As anticipated we find that $\Delta \epsilon(\bk)$ grows logarithmically, and thus that the absence of screening or a UV cutoff is associated with logarithmically divergent corrections to the theory. 
 
\begin{figure}[!htb]
\includegraphics[width=\linewidth]{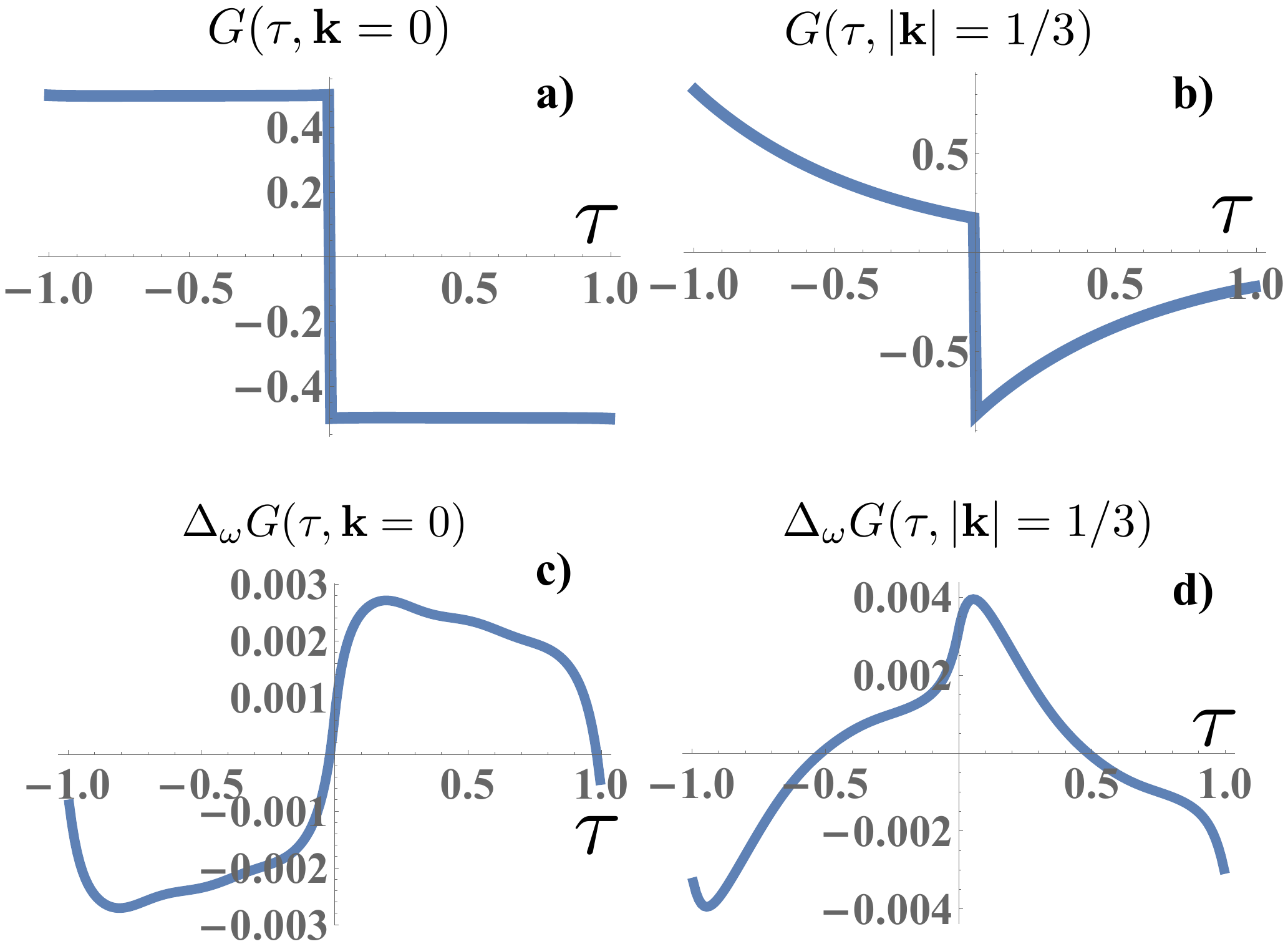}
\caption{
Greens function of the interacting system at the nodal point (a) and at $|\bk|=1/3$ (b). The displayed eigenvalue corresponds to positive energy. The corrections (Eq. \ref{dGo}) from the frequency dependent part of the self energy (c-d) are sub-leading by three orders of magnitude.  
The data corresponds to an expansion order $O=3$, while the model parameters are given by $\Lambda_0=4\pi,\;\lambda^{-1}_0=1/4, \alpha=2, \;v_f^0=4,\; T=1$, i.e. correspond to $T=1$ in Fig. (\ref{Sigma_0},a).
}
\label{G1}
\end{figure}

\begin{figure}[!htb]
\includegraphics[width=\linewidth]{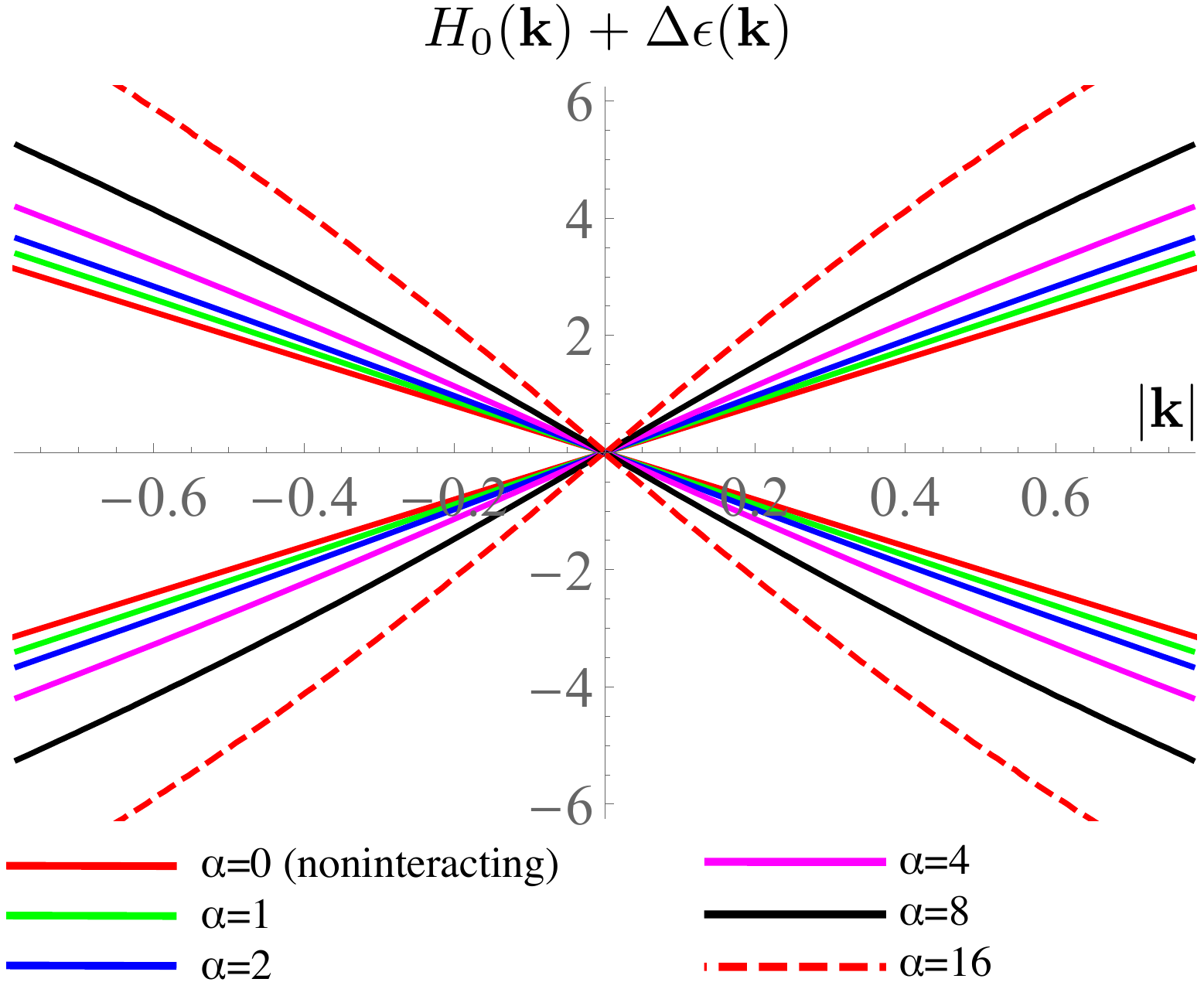}
\caption{
Dispersion with frequency-independent corrections, at interaction strengths ranging from $\alpha=0$, i.e. noninteracting, to $\alpha=16$. As the contribution from the frequency-dependent part of the self energy is small for the parameter values that we consider (see Fig. \ref{G1}), 
this represents a good approximation of the overall electronic structure.
The parameter values are given by $T=1/8, \; v_f^0=4,\;\Lambda_0=\pi,\; \lambda_0^{-1}=1/4$.
}
\label{cone}
\end{figure}

 Finally we consider the frequency dependent part of the self energy. In principle we can define a Greens function that takes into account the corrections to the single particle energy according to
 \bea
G'_0(\omega,\bk)=\frac{1}{i\omega-H_0(\bk)-\Delta \epsilon (\bk)},\label{G0prim}
\eea
so that the frequency dependent contribution takes the form
\bea
\Delta_\omega  G(\omega,\bk)=G(\omega,\bk)-G'_0(\omega,\bk)\label{dGo}.
\eea 
This term then encodes corrections that are not reconcilable with a free-fermion description, notably that a given momentum is associated with a {\it spectrum} of quasiparticles. 
In Fig. (\ref{G1}) the full Greens function is displayed, together with the frequency dependent correction (Eq. \ref{dGo}) at the nodal point and also at finite momentum. In both cases we find that the second part is sub-leading by $\sim3$ orders of magnitude. This can most likely be attributed to the linear dispersion: With a rapidly vanishing density of states near the Fermi level, processes that involve excitations of the environment are dramatically suppressed compared to in a conventional metal or indeed even a semimetal with quadratic dispersion.
Thus, for the parameters considered, a description in terms of free fermions with a renormalized dispersion appears to accurately capture the electronic structure of the interacting system. In this approximation the corrections to the Fermi velocity are given by Fig. (\ref{Sigma_0}). Likewise, $H_0(\bk)+\Sigma(\omega=0,\bk)$, provides a renormalized band-theory description in the presence of screened Coulomb interactions, see Fig. (\ref{cone}).
It should be stressed that corrections to the dispersion are by no means small and diverge according to Eq. (\ref{conj2}).

The generalization of these results to realistic models of real materials hinges on two principal considerations: Firstly, as alluded to in the introduction, electrons that exist on a lattice can in principle form Mott insulating states for sufficiently strong interactions. Secondly, Weyl nodes must necessarily appear in pairs, meaning that the self energy will include contributions from regions in $\bk$-space where the dispersion is not well described by Eq. (\ref{H0}). The principal importance of these terms is strongly model dependent, and decreases sharply with increasing screening length, and to a certain extent also with larger node separation, see supplementary material for a discussion, \cite{Supplemental}.
In this context it is interesting to note the prediction, that nodal points with a separation that is of the same order of magnitude as the Brilluoin may be realized in the magnetic Heusler alloys \cite{PhysRevLett.117.236401}.

%%%%%%%%%% CONCLUSION %%%%%%%%%%%%%%%%%
{\it Conclusions.---}
We have examined the premises for the stability of the semimetallic phase and convergence of the diagrammatic expansion in the case of a single node in a Weyl semimetal. In the perturbative regime where the series converges, we note that nodal points are in principle protected by the symmetry of the self energy \cite{PhysRevB.97.161102}. Using a scaling transformation we find that the zero-temperature limit can be equated to the simultaneous divergence of the ultraviolet cutoff and the inverse screening. Correspondingly, the ground state is characterized by the ratio of the two, with logarithmically divergent corrections in the limit $\Lambda_0/\lambda_0^{-1}\to\infty$. For finite ratios we do however find that the diagrammatic expansion remains convergent in the low temperature regime, and that the corrections to the dispersion become temperature independent.
Thus, Coulomb interactions in the Weyl semimetals play a similar role as in graphene, where corrections to the Fermi velocity diverge logarithmically with system size at zero temperature \cite{PhysRevLett.118.026403}.

While the ultraviolet cutoff in real world materials is essentially determined by lattice spacing, screening is greatly affected by parasitic bands, which suggests that these can have a dramatic effect on the Fermi velocity of an interacting system. Likewise, as Coulomb repulsion generally reduces the tilt \cite{PhysRevB.96.195157}, screening may be extremely important in this respect as well. Also we note that for renormalisation group calculations the relevant parameter is in principle the ratio $\Lambda_0/\lambda_0^{-1}$.

We find that the convergence is rapid with respect to expansion order, and that most of the correction to the dispersion is contained in the Fock term. Meanwhile, the contribution to the Greens function coming from the frequency-dependent part of the self energy is extremely small, as should be expected in a Dirac system. Correspondingly we arrive at a paradigm of effectively almost free fermions, though with large corrections to the dispersion.

\acknowledgements
{\it Acknowledgements.---} 
This work was supported by the Swedish research council (VR) and the Wallenberg Academy Fellows program of the Knut and Alice Wallenberg Foundation. Computations were performed on resources provided by the Swedish National Infrastructure for Computing (SNIC) at the High Performance Computing Center North in Umeå, Sweden.

\bibliography{biblio}

\begin{thebibliography}{27}%
\makeatletter
\providecommand \@ifxundefined [1]{%
 \@ifx{#1\undefined}
}%
\providecommand \@ifnum [1]{%
 \ifnum #1\expandafter \@firstoftwo
 \else \expandafter \@secondoftwo
 \fi
}%
\providecommand \@ifx [1]{%
 \ifx #1\expandafter \@firstoftwo
 \else \expandafter \@secondoftwo
 \fi
}%
\providecommand \natexlab [1]{#1}%
\providecommand \enquote  [1]{``#1''}%
\providecommand \bibnamefont  [1]{#1}%
\providecommand \bibfnamefont [1]{#1}%
\providecommand \citenamefont [1]{#1}%
\providecommand \href@noop [0]{\@secondoftwo}%
\providecommand \href [0]{\begingroup \@sanitize@url \@href}%
\providecommand \@href[1]{\@@startlink{#1}\@@href}%
\providecommand \@@href[1]{\endgroup#1\@@endlink}%
\providecommand \@sanitize@url [0]{\catcode `\\12\catcode `\$12\catcode
  `\&12\catcode `\#12\catcode `\^12\catcode `\_12\catcode `\%12\relax}%
\providecommand \@@startlink[1]{}%
\providecommand \@@endlink[0]{}%
\providecommand \url  [0]{\begingroup\@sanitize@url \@url }%
\providecommand \@url [1]{\endgroup\@href {#1}{\urlprefix }}%
\providecommand \urlprefix  [0]{URL }%
\providecommand \Eprint [0]{\href }%
\providecommand \doibase [0]{http://dx.doi.org/}%
\providecommand \selectlanguage [0]{\@gobble}%
\providecommand \bibinfo  [0]{\@secondoftwo}%
\providecommand \bibfield  [0]{\@secondoftwo}%
\providecommand \translation [1]{[#1]}%
\providecommand \BibitemOpen [0]{}%
\providecommand \bibitemStop [0]{}%
\providecommand \bibitemNoStop [0]{.\EOS\space}%
\providecommand \EOS [0]{\spacefactor3000\relax}%
\providecommand \BibitemShut  [1]{\csname bibitem#1\endcsname}%
\let\auto@bib@innerbib\@empty
%</preamble>
\bibitem [{\citenamefont {Armitage}\ \emph {et~al.}(2018)\citenamefont
  {Armitage}, \citenamefont {Mele},\ and\ \citenamefont
  {Vishwanath}}]{RevModPhys.90.015001}%
  \BibitemOpen
  \bibfield  {author} {\bibinfo {author} {\bibfnamefont {N.~P.}\ \bibnamefont
  {Armitage}}, \bibinfo {author} {\bibfnamefont {E.~J.}\ \bibnamefont {Mele}},
  \ and\ \bibinfo {author} {\bibfnamefont {A.}~\bibnamefont {Vishwanath}},\
  }\href {\doibase 10.1103/RevModPhys.90.015001} {\bibfield  {journal}
  {\bibinfo  {journal} {Rev. Mod. Phys.}\ }\textbf {\bibinfo {volume} {90}},\
  \bibinfo {pages} {015001} (\bibinfo {year} {2018})}\BibitemShut {NoStop}%
\bibitem [{\citenamefont {Volovik}(2009)}]{volovik2009universe}%
  \BibitemOpen
  \bibfield  {author} {\bibinfo {author} {\bibfnamefont {G.}~\bibnamefont
  {Volovik}},\ }\href {https://books.google.se/books?id=6uj76kFJOHEC} {\emph
  {\bibinfo {title} {The Universe in a Helium Droplet}}},\ International Series
  of Monographs on Physics\ (\bibinfo  {publisher} {OUP Oxford},\ \bibinfo
  {year} {2009})\BibitemShut {NoStop}%
\bibitem [{\citenamefont {Wan}\ \emph {et~al.}(2011)\citenamefont {Wan},
  \citenamefont {Turner}, \citenamefont {Vishwanath},\ and\ \citenamefont
  {Savrasov}}]{PhysRevB.83.205101}%
  \BibitemOpen
  \bibfield  {author} {\bibinfo {author} {\bibfnamefont {X.}~\bibnamefont
  {Wan}}, \bibinfo {author} {\bibfnamefont {A.~M.}\ \bibnamefont {Turner}},
  \bibinfo {author} {\bibfnamefont {A.}~\bibnamefont {Vishwanath}}, \ and\
  \bibinfo {author} {\bibfnamefont {S.~Y.}\ \bibnamefont {Savrasov}},\ }\href
  {\doibase 10.1103/PhysRevB.83.205101} {\bibfield  {journal} {\bibinfo
  {journal} {Phys. Rev. B}\ }\textbf {\bibinfo {volume} {83}},\ \bibinfo
  {pages} {205101} (\bibinfo {year} {2011})}\BibitemShut {NoStop}%
\bibitem [{\citenamefont {Murakami}(2007)}]{1367-2630-9-9-356}%
  \BibitemOpen
  \bibfield  {author} {\bibinfo {author} {\bibfnamefont {S.}~\bibnamefont
  {Murakami}},\ }\href {http://stacks.iop.org/1367-2630/9/i=9/a=356} {\bibfield
   {journal} {\bibinfo  {journal} {New Journal of Physics}\ }\textbf {\bibinfo
  {volume} {9}},\ \bibinfo {pages} {356} (\bibinfo {year} {2007})}\BibitemShut
  {NoStop}%
\bibitem [{\citenamefont {Hosur}\ and\ \citenamefont
  {Qi}(2013)}]{HOSUR2013857}%
  \BibitemOpen
  \bibfield  {author} {\bibinfo {author} {\bibfnamefont {P.}~\bibnamefont
  {Hosur}}\ and\ \bibinfo {author} {\bibfnamefont {X.}~\bibnamefont {Qi}},\
  }\href
  {http://www.sciencedirect.com/science/article/pii/S1631070513001710?via%3Dihub}
  {\bibfield  {journal} {\bibinfo  {journal} {Comptes Rendus Physique}\
  }\textbf {\bibinfo {volume} {14}},\ \bibinfo {pages} {857 } (\bibinfo {year}
  {2013})},\ \bibinfo {note} {topological insulators / Isolants
  topologiques}\BibitemShut {NoStop}%
\bibitem [{\citenamefont {Hasan}\ \emph {et~al.}(2017)\citenamefont {Hasan},
  \citenamefont {Xu}, \citenamefont {Belopolski},\ and\ \citenamefont
  {Huang}}]{doi:10.1146/annurev-conmatphys-031016-025225}%
  \BibitemOpen
  \bibfield  {author} {\bibinfo {author} {\bibfnamefont {M.~Z.}\ \bibnamefont
  {Hasan}}, \bibinfo {author} {\bibfnamefont {S.-Y.}\ \bibnamefont {Xu}},
  \bibinfo {author} {\bibfnamefont {I.}~\bibnamefont {Belopolski}}, \ and\
  \bibinfo {author} {\bibfnamefont {S.-M.}\ \bibnamefont {Huang}},\ }\href
  {\doibase 10.1146/annurev-conmatphys-031016-025225} {\bibfield  {journal}
  {\bibinfo  {journal} {Annual Review of Condensed Matter Physics}\ }\textbf
  {\bibinfo {volume} {8}},\ \bibinfo {pages} {289} (\bibinfo {year}
  {2017})}\BibitemShut {NoStop}%
\bibitem [{\citenamefont {Weng}\ \emph {et~al.}(2015)\citenamefont {Weng},
  \citenamefont {Fang}, \citenamefont {Fang}, \citenamefont {Bernevig},\ and\
  \citenamefont {Dai}}]{PhysRevX.5.011029}%
  \BibitemOpen
  \bibfield  {author} {\bibinfo {author} {\bibfnamefont {H.}~\bibnamefont
  {Weng}}, \bibinfo {author} {\bibfnamefont {C.}~\bibnamefont {Fang}}, \bibinfo
  {author} {\bibfnamefont {Z.}~\bibnamefont {Fang}}, \bibinfo {author}
  {\bibfnamefont {B.~A.}\ \bibnamefont {Bernevig}}, \ and\ \bibinfo {author}
  {\bibfnamefont {X.}~\bibnamefont {Dai}},\ }\href {\doibase
  10.1103/PhysRevX.5.011029} {\bibfield  {journal} {\bibinfo  {journal} {Phys.
  Rev. X}\ }\textbf {\bibinfo {volume} {5}},\ \bibinfo {pages} {011029}
  (\bibinfo {year} {2015})}\BibitemShut {NoStop}%
\bibitem [{\citenamefont {Burkov}\ \emph {et~al.}(2011)\citenamefont {Burkov},
  \citenamefont {Hook},\ and\ \citenamefont {Balents}}]{PhysRevB.84.235126}%
  \BibitemOpen
  \bibfield  {author} {\bibinfo {author} {\bibfnamefont {A.~A.}\ \bibnamefont
  {Burkov}}, \bibinfo {author} {\bibfnamefont {M.~D.}\ \bibnamefont {Hook}}, \
  and\ \bibinfo {author} {\bibfnamefont {L.}~\bibnamefont {Balents}},\ }\href
  {\doibase 10.1103/PhysRevB.84.235126} {\bibfield  {journal} {\bibinfo
  {journal} {Phys. Rev. B}\ }\textbf {\bibinfo {volume} {84}},\ \bibinfo
  {pages} {235126} (\bibinfo {year} {2011})}\BibitemShut {NoStop}%
\bibitem [{\citenamefont {Xu}\ \emph {et~al.}(2015)\citenamefont {Xu},
  \citenamefont {Belopolski}, \citenamefont {Alidoust}, \citenamefont
  {Neupane}, \citenamefont {Bian}, \citenamefont {Zhang}, \citenamefont
  {Sankar}, \citenamefont {Chang}, \citenamefont {Yuan}, \citenamefont {Lee},
  \citenamefont {Huang}, \citenamefont {Zheng}, \citenamefont {Ma},
  \citenamefont {Sanchez}, \citenamefont {Wang}, \citenamefont {Bansil},
  \citenamefont {Chou}, \citenamefont {Shibayev}, \citenamefont {Lin},
  \citenamefont {Jia},\ and\ \citenamefont {Hasan}}]{Xu613}%
  \BibitemOpen
  \bibfield  {author} {\bibinfo {author} {\bibfnamefont {S.-Y.}\ \bibnamefont
  {Xu}}, \bibinfo {author} {\bibfnamefont {I.}~\bibnamefont {Belopolski}},
  \bibinfo {author} {\bibfnamefont {N.}~\bibnamefont {Alidoust}}, \bibinfo
  {author} {\bibfnamefont {M.}~\bibnamefont {Neupane}}, \bibinfo {author}
  {\bibfnamefont {G.}~\bibnamefont {Bian}}, \bibinfo {author} {\bibfnamefont
  {C.}~\bibnamefont {Zhang}}, \bibinfo {author} {\bibfnamefont
  {R.}~\bibnamefont {Sankar}}, \bibinfo {author} {\bibfnamefont
  {G.}~\bibnamefont {Chang}}, \bibinfo {author} {\bibfnamefont
  {Z.}~\bibnamefont {Yuan}}, \bibinfo {author} {\bibfnamefont {C.-C.}\
  \bibnamefont {Lee}}, \bibinfo {author} {\bibfnamefont {S.-M.}\ \bibnamefont
  {Huang}}, \bibinfo {author} {\bibfnamefont {H.}~\bibnamefont {Zheng}},
  \bibinfo {author} {\bibfnamefont {J.}~\bibnamefont {Ma}}, \bibinfo {author}
  {\bibfnamefont {D.~S.}\ \bibnamefont {Sanchez}}, \bibinfo {author}
  {\bibfnamefont {B.}~\bibnamefont {Wang}}, \bibinfo {author} {\bibfnamefont
  {A.}~\bibnamefont {Bansil}}, \bibinfo {author} {\bibfnamefont
  {F.}~\bibnamefont {Chou}}, \bibinfo {author} {\bibfnamefont {P.~P.}\
  \bibnamefont {Shibayev}}, \bibinfo {author} {\bibfnamefont {H.}~\bibnamefont
  {Lin}}, \bibinfo {author} {\bibfnamefont {S.}~\bibnamefont {Jia}}, \ and\
  \bibinfo {author} {\bibfnamefont {M.~Z.}\ \bibnamefont {Hasan}},\ }\href
  {\doibase 10.1126/science.aaa9297} {\bibfield  {journal} {\bibinfo  {journal}
  {Science}\ }\textbf {\bibinfo {volume} {349}},\ \bibinfo {pages} {613}
  (\bibinfo {year} {2015})},\ \Eprint
  {http://arxiv.org/abs/http://science.sciencemag.org/content/349/6248/613.full.pdf}
  {http://science.sciencemag.org/content/349/6248/613.full.pdf} \BibitemShut
  {NoStop}%
\bibitem [{\citenamefont {Trescher}\ \emph {et~al.}(2015)\citenamefont
  {Trescher}, \citenamefont {Sbierski}, \citenamefont {Brouwer},\ and\
  \citenamefont {Bergholtz}}]{PhysRevB.91.115135}%
  \BibitemOpen
  \bibfield  {author} {\bibinfo {author} {\bibfnamefont {M.}~\bibnamefont
  {Trescher}}, \bibinfo {author} {\bibfnamefont {B.}~\bibnamefont {Sbierski}},
  \bibinfo {author} {\bibfnamefont {P.~W.}\ \bibnamefont {Brouwer}}, \ and\
  \bibinfo {author} {\bibfnamefont {E.~J.}\ \bibnamefont {Bergholtz}},\ }\href
  {\doibase 10.1103/PhysRevB.91.115135} {\bibfield  {journal} {\bibinfo
  {journal} {Phys. Rev. B}\ }\textbf {\bibinfo {volume} {91}},\ \bibinfo
  {pages} {115135} (\bibinfo {year} {2015})}\BibitemShut {NoStop}%
\bibitem [{\citenamefont {Soluyanov}\ \emph {et~al.}(2015)\citenamefont
  {Soluyanov}, \citenamefont {Gresch}, \citenamefont {Wang}, \citenamefont
  {Wu}, \citenamefont {Troyer}, \citenamefont {Dai},\ and\ \citenamefont
  {Bernevig}}]{10.1038/nature15768}%
  \BibitemOpen
  \bibfield  {author} {\bibinfo {author} {\bibfnamefont {A.~A.}\ \bibnamefont
  {Soluyanov}}, \bibinfo {author} {\bibfnamefont {D.}~\bibnamefont {Gresch}},
  \bibinfo {author} {\bibfnamefont {Z.}~\bibnamefont {Wang}}, \bibinfo {author}
  {\bibfnamefont {Q.}~\bibnamefont {Wu}}, \bibinfo {author} {\bibfnamefont
  {M.}~\bibnamefont {Troyer}}, \bibinfo {author} {\bibfnamefont
  {X.}~\bibnamefont {Dai}}, \ and\ \bibinfo {author} {\bibfnamefont {B.~A.}\
  \bibnamefont {Bernevig}},\ }\href {\doibase 10.1038/nature15768} {\bibfield
  {journal} {\bibinfo  {journal} {Nature}\ }\textbf {\bibinfo {volume} {527}},\
  \bibinfo {pages} {495} (\bibinfo {year} {2015})}\BibitemShut {NoStop}%
\bibitem [{\citenamefont {Detassis}\ \emph {et~al.}(2017)\citenamefont
  {Detassis}, \citenamefont {Fritz},\ and\ \citenamefont
  {Grubinskas}}]{PhysRevB.96.195157}%
  \BibitemOpen
  \bibfield  {author} {\bibinfo {author} {\bibfnamefont {F.}~\bibnamefont
  {Detassis}}, \bibinfo {author} {\bibfnamefont {L.}~\bibnamefont {Fritz}}, \
  and\ \bibinfo {author} {\bibfnamefont {S.}~\bibnamefont {Grubinskas}},\
  }\href {\doibase 10.1103/PhysRevB.96.195157} {\bibfield  {journal} {\bibinfo
  {journal} {Phys. Rev. B}\ }\textbf {\bibinfo {volume} {96}},\ \bibinfo
  {pages} {195157} (\bibinfo {year} {2017})}\BibitemShut {NoStop}%
\bibitem [{\citenamefont {Witczak-Krempa}\ \emph {et~al.}(2014)\citenamefont
  {Witczak-Krempa}, \citenamefont {Knap},\ and\ \citenamefont
  {Abanin}}]{PhysRevLett.113.136402}%
  \BibitemOpen
  \bibfield  {author} {\bibinfo {author} {\bibfnamefont {W.}~\bibnamefont
  {Witczak-Krempa}}, \bibinfo {author} {\bibfnamefont {M.}~\bibnamefont
  {Knap}}, \ and\ \bibinfo {author} {\bibfnamefont {D.}~\bibnamefont
  {Abanin}},\ }\href {\doibase 10.1103/PhysRevLett.113.136402} {\bibfield
  {journal} {\bibinfo  {journal} {Phys. Rev. Lett.}\ }\textbf {\bibinfo
  {volume} {113}},\ \bibinfo {pages} {136402} (\bibinfo {year}
  {2014})}\BibitemShut {NoStop}%
\bibitem [{\citenamefont {Morimoto}\ and\ \citenamefont
  {Nagaosa}(2016)}]{10.1038/srep19853}%
  \BibitemOpen
  \bibfield  {author} {\bibinfo {author} {\bibfnamefont {T.}~\bibnamefont
  {Morimoto}}\ and\ \bibinfo {author} {\bibfnamefont {N.}~\bibnamefont
  {Nagaosa}},\ }\href {\doibase 10.1038/srep19853} {\bibfield  {journal}
  {\bibinfo  {journal} {Scientific Reports}\ }\textbf {\bibinfo {volume} {6}},\
  \bibinfo {pages} {19853} (\bibinfo {year} {2016})}\BibitemShut {NoStop}%
\bibitem [{\citenamefont {{Meng}}\ and\ \citenamefont
  {{Budich}}(2018)}]{2018arXiv180405078M}%
  \BibitemOpen
  \bibfield  {author} {\bibinfo {author} {\bibfnamefont {T.}~\bibnamefont
  {{Meng}}}\ and\ \bibinfo {author} {\bibfnamefont {J.~C.}\ \bibnamefont
  {{Budich}}},\ }\href@noop {} {\bibfield  {journal} {\bibinfo  {journal}
  {ArXiv e-prints}\ } (\bibinfo {year} {2018})},\ \Eprint
  {http://arxiv.org/abs/1804.05078} {arXiv:1804.05078 [cond-mat.str-el]}
  \BibitemShut {NoStop}%
\bibitem [{\citenamefont {Araki}(2015)}]{Araki}%
  \BibitemOpen
  \bibfield  {author} {\bibinfo {author} {\bibfnamefont {Y.}~\bibnamefont
  {Araki}},\ }\href {https://pos.sissa.it/251/046/pdf} {\bibfield  {journal}
  {\bibinfo  {journal} {Proceedings of Science, Lattice 2015}\ }\textbf
  {\bibinfo {volume} {046}} (\bibinfo {year} {2015})}\BibitemShut {NoStop}%
\bibitem [{\citenamefont {Laubach}\ \emph {et~al.}(2016)\citenamefont
  {Laubach}, \citenamefont {Platt}, \citenamefont {Thomale}, \citenamefont
  {Neupert},\ and\ \citenamefont {Rachel}}]{PhysRevB.94.241102}%
  \BibitemOpen
  \bibfield  {author} {\bibinfo {author} {\bibfnamefont {M.}~\bibnamefont
  {Laubach}}, \bibinfo {author} {\bibfnamefont {C.}~\bibnamefont {Platt}},
  \bibinfo {author} {\bibfnamefont {R.}~\bibnamefont {Thomale}}, \bibinfo
  {author} {\bibfnamefont {T.}~\bibnamefont {Neupert}}, \ and\ \bibinfo
  {author} {\bibfnamefont {S.}~\bibnamefont {Rachel}},\ }\href {\doibase
  10.1103/PhysRevB.94.241102} {\bibfield  {journal} {\bibinfo  {journal} {Phys.
  Rev. B}\ }\textbf {\bibinfo {volume} {94}},\ \bibinfo {pages} {241102}
  (\bibinfo {year} {2016})}\BibitemShut {NoStop}%
\bibitem [{\citenamefont {Hofmann}\ \emph {et~al.}(2015)\citenamefont
  {Hofmann}, \citenamefont {Barnes},\ and\ \citenamefont
  {Das~Sarma}}]{PhysRevB.92.045104}%
  \BibitemOpen
  \bibfield  {author} {\bibinfo {author} {\bibfnamefont {J.}~\bibnamefont
  {Hofmann}}, \bibinfo {author} {\bibfnamefont {E.}~\bibnamefont {Barnes}}, \
  and\ \bibinfo {author} {\bibfnamefont {S.}~\bibnamefont {Das~Sarma}},\ }\href
  {\doibase 10.1103/PhysRevB.92.045104} {\bibfield  {journal} {\bibinfo
  {journal} {Phys. Rev. B}\ }\textbf {\bibinfo {volume} {92}},\ \bibinfo
  {pages} {045104} (\bibinfo {year} {2015})}\BibitemShut {NoStop}%
\bibitem [{\citenamefont {Carlstr\"om}\ and\ \citenamefont
  {Bergholtz}(2018)}]{PhysRevB.97.161102}%
  \BibitemOpen
  \bibfield  {author} {\bibinfo {author} {\bibfnamefont {J.}~\bibnamefont
  {Carlstr\"om}}\ and\ \bibinfo {author} {\bibfnamefont {E.~J.}\ \bibnamefont
  {Bergholtz}},\ }\href {\doibase 10.1103/PhysRevB.97.161102} {\bibfield
  {journal} {\bibinfo  {journal} {Phys. Rev. B}\ }\textbf {\bibinfo {volume}
  {97}},\ \bibinfo {pages} {161102} (\bibinfo {year} {2018})}\BibitemShut
  {NoStop}%
\bibitem [{\citenamefont {Giuliani}\ and\ \citenamefont
  {Mastropietro}(2009)}]{PhysRevB.79.201403}%
  \BibitemOpen
  \bibfield  {author} {\bibinfo {author} {\bibfnamefont {A.}~\bibnamefont
  {Giuliani}}\ and\ \bibinfo {author} {\bibfnamefont {V.}~\bibnamefont
  {Mastropietro}},\ }\href {\doibase 10.1103/PhysRevB.79.201403} {\bibfield
  {journal} {\bibinfo  {journal} {Phys. Rev. B}\ }\textbf {\bibinfo {volume}
  {79}},\ \bibinfo {pages} {201403} (\bibinfo {year} {2009})}\BibitemShut
  {NoStop}%
\bibitem [{\citenamefont {Tupitsyn}\ and\ \citenamefont
  {Prokof'ev}(2017)}]{PhysRevLett.118.026403}%
  \BibitemOpen
  \bibfield  {author} {\bibinfo {author} {\bibfnamefont {I.~S.}\ \bibnamefont
  {Tupitsyn}}\ and\ \bibinfo {author} {\bibfnamefont {N.~V.}\ \bibnamefont
  {Prokof'ev}},\ }\href {\doibase 10.1103/PhysRevLett.118.026403} {\bibfield
  {journal} {\bibinfo  {journal} {Phys. Rev. Lett.}\ }\textbf {\bibinfo
  {volume} {118}},\ \bibinfo {pages} {026403} (\bibinfo {year}
  {2017})}\BibitemShut {NoStop}%
\bibitem [{\citenamefont {Kotov}\ \emph {et~al.}(2012)\citenamefont {Kotov},
  \citenamefont {Uchoa}, \citenamefont {Pereira}, \citenamefont {Guinea},\ and\
  \citenamefont {Castro~Neto}}]{RevModPhys.84.1067}%
  \BibitemOpen
  \bibfield  {author} {\bibinfo {author} {\bibfnamefont {V.~N.}\ \bibnamefont
  {Kotov}}, \bibinfo {author} {\bibfnamefont {B.}~\bibnamefont {Uchoa}},
  \bibinfo {author} {\bibfnamefont {V.~M.}\ \bibnamefont {Pereira}}, \bibinfo
  {author} {\bibfnamefont {F.}~\bibnamefont {Guinea}}, \ and\ \bibinfo {author}
  {\bibfnamefont {A.~H.}\ \bibnamefont {Castro~Neto}},\ }\href {\doibase
  10.1103/RevModPhys.84.1067} {\bibfield  {journal} {\bibinfo  {journal} {Rev.
  Mod. Phys.}\ }\textbf {\bibinfo {volume} {84}},\ \bibinfo {pages} {1067}
  (\bibinfo {year} {2012})}\BibitemShut {NoStop}%
\bibitem [{\citenamefont {Fetter}\ and\ \citenamefont
  {Walecka}(1971)}]{fetter}%
  \BibitemOpen
  \bibfield  {author} {\bibinfo {author} {\bibfnamefont {A.~L.}\ \bibnamefont
  {Fetter}}\ and\ \bibinfo {author} {\bibfnamefont {J.~D.}\ \bibnamefont
  {Walecka}},\ }\href@noop {} {\emph {\bibinfo {title} {Quantum theory of
  many-particle systems}}}\ (\bibinfo  {publisher} {Dover Publications},\
  \bibinfo {year} {1971})\BibitemShut {NoStop}%
\bibitem [{\citenamefont {Prokof'ev}\ and\ \citenamefont
  {Svistunov}(2007)}]{PhysRevLett.99.250201}%
  \BibitemOpen
  \bibfield  {author} {\bibinfo {author} {\bibfnamefont {N.}~\bibnamefont
  {Prokof'ev}}\ and\ \bibinfo {author} {\bibfnamefont {B.}~\bibnamefont
  {Svistunov}},\ }\href {\doibase 10.1103/PhysRevLett.99.250201} {\bibfield
  {journal} {\bibinfo  {journal} {Phys. Rev. Lett.}\ }\textbf {\bibinfo
  {volume} {99}},\ \bibinfo {pages} {250201} (\bibinfo {year}
  {2007})}\BibitemShut {NoStop}%
\bibitem [{\citenamefont {Carlstr\"om}(2018)}]{PhysRevB.97.075119}%
  \BibitemOpen
  \bibfield  {author} {\bibinfo {author} {\bibfnamefont {J.}~\bibnamefont
  {Carlstr\"om}},\ }\href {\doibase 10.1103/PhysRevB.97.075119} {\bibfield
  {journal} {\bibinfo  {journal} {Phys. Rev. B}\ }\textbf {\bibinfo {volume}
  {97}},\ \bibinfo {pages} {075119} (\bibinfo {year} {2018})}\BibitemShut
  {NoStop}%
\bibitem [{Sup()}]{Supplemental}%
  \BibitemOpen
  \href@noop {} {\ }\bibinfo {note} {See supplemental material at [url] for a
  discussion about the generalisation of results for the linearised model to
  full lattice models}\BibitemShut {NoStop}%
\bibitem [{\citenamefont {Wang}\ \emph {et~al.}(2016)\citenamefont {Wang},
  \citenamefont {Vergniory}, \citenamefont {Kushwaha}, \citenamefont
  {Hirschberger}, \citenamefont {Chulkov}, \citenamefont {Ernst}, \citenamefont
  {Ong}, \citenamefont {Cava},\ and\ \citenamefont
  {Bernevig}}]{PhysRevLett.117.236401}%
  \BibitemOpen
  \bibfield  {author} {\bibinfo {author} {\bibfnamefont {Z.}~\bibnamefont
  {Wang}}, \bibinfo {author} {\bibfnamefont {M.~G.}\ \bibnamefont {Vergniory}},
  \bibinfo {author} {\bibfnamefont {S.}~\bibnamefont {Kushwaha}}, \bibinfo
  {author} {\bibfnamefont {M.}~\bibnamefont {Hirschberger}}, \bibinfo {author}
  {\bibfnamefont {E.~V.}\ \bibnamefont {Chulkov}}, \bibinfo {author}
  {\bibfnamefont {A.}~\bibnamefont {Ernst}}, \bibinfo {author} {\bibfnamefont
  {N.~P.}\ \bibnamefont {Ong}}, \bibinfo {author} {\bibfnamefont {R.~J.}\
  \bibnamefont {Cava}}, \ and\ \bibinfo {author} {\bibfnamefont {B.~A.}\
  \bibnamefont {Bernevig}},\ }\href {\doibase 10.1103/PhysRevLett.117.236401}
  {\bibfield  {journal} {\bibinfo  {journal} {Phys. Rev. Lett.}\ }\textbf
  {\bibinfo {volume} {117}},\ \bibinfo {pages} {236401} (\bibinfo {year}
  {2016})}\BibitemShut {NoStop}%
\end{thebibliography}%

\end{document}